\documentstyle[11pt,paspconf,epsf]{article}

\begin{document}
%----------------------------------------------------------------------
\title{Abundance profile and stellar content of IZw18 }
\author{Fran\c cois Legrand}
\affil{Institut d'Astrophysique de Paris, 98bis Bd Arago, 75014 Paris, France}

%----------------------------------------------------------------------
\begin{abstract}
New observations of the metal poor galaxy IZw 18 are discussed.  
Wolf-Rayet stars of WC type have been detected
 in the NW-HII region contrary to evolutionary synthesis model predictions.
Implications on the mass loss rate and on the 
formation processes of WR stars are discussed. A very homogeneous metal 
abundance is observed within the HII region.  This emphasizes the problem of the dispersal and 
mixing of new synthesized element in a starburst. Different scenarios are 
discussed, showing that metals 
remain most likely hidden in a hot phase and that the observed present metallicity is the 
result of a previous star formation event.
\end{abstract}

%----------------------------------------------------------------------
% KEYWORDS SHOULD BE INCLUDED, BUT THEY ARE NOT PRINTED IN THE HARDCOPY!
\keywords{Starburst Galaxies -- Mixing -- IZw 18 -- Wolf-Rayet stars -- Abundances}

%----------------------------------------------------------------------
\section{Introduction}
The blue compact galaxy IZw 18 is still the lowest metallicity galaxy known
experiencing an intense star formation episode. Moreover, the metallicity
in the HI halo seems to be lower than in the HII region (Kunth et al, 1994).
This makes IZw 18 a good candidate to study the processes by which  massive 
stars enrich the interstellar medium in a starburst galaxy.

%----------------------------------------------------------------------
\section{Observations}
%----------------------------------
Deep long-slit observations (14 hours with a thick CCD) of the NW region of 
IZw 18 were made in 1995 February at the CFH Telescope. The position angle was 45$\deg$, 
covering a spectral range from 3700 to 6900 $\rm \AA$. 
The spatial and spectral resolution were 0.3145 arcsec/pix and  $\rm 8.2 \ \AA $
respectively. 
The seeing was between 1 and 1.5 arcsec. In order to measure the temperature
sensitive line OIII[4363] with a good accuracy, the spectrum was 
binned over 1.6''  at all the positions along the slit. This allows us
to measure the temperature and derive abundances over more than 600 pc. 
Two major results were obtained as discussed below.

\section{Stellar content and abundance of IZw 18}

First, we found WR stars of the WC type in IZw 18 (Legrand et al, 1997). 
The presence of this kind of stars
was unexpected in such a low metallicity object because evolutionary synthesis
 models predict few WR stars of only WN types. This should indicate that
 massive stars with mass larger than $80 M_{\odot}$ are present and that 
mass loss rates may be larger than twice the standard one at low 
metallicity. 
Alternatively, the low number of WC detected (a few) is also  
compatible with a binary channel of formation without requiring high 
mass loss rate.
However, a larger number of WR stars (5 WC and 17 WN) have been detected in 
IZw 18 at another slit position (Izotov et al, 1997). If this is confirmed, 
the binary channel for WR star formation would not be sufficient to account
for such a large number and high mass loss rate would also be required.
Perhaps more importantly, we found a strong correlation between the spatial location of the
WRs and the maximum of the nebular emission line HeII[4686], suggesting that
 WC stars should indeed be at the origin of this line, as proposed by 
Schaerer (1996). 

\begin{figure}
% psfile=#1 vsize=#2 angle=#3 hscale=#4 vscale=#5 hoffset=#6 voffset=#7
\plotfiddle{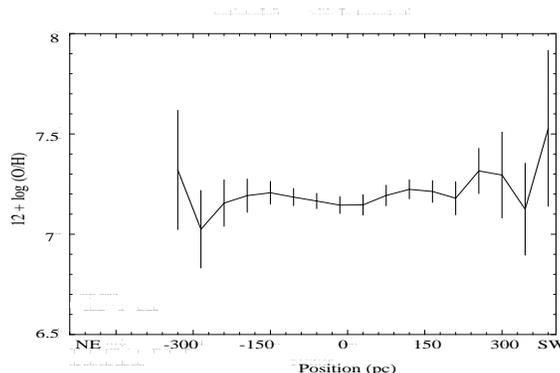}{36truemm}{-90}{40}{26}{-150}{129}
\caption{Spatial abundance profile in IZw 18.}
\label{abundance}
\end{figure}

The second result (Legrand et al, A\&A in preparation) is that the oxygen 
abundance does not present significant variation at scales smaller than 
600 pc (Fig. \ref{abundance}). As the spatial resolution of our observations 
is 50 pc, smaller scale inhomogeneities cannot be excluded.
This is comparable with results obtained for other starburst
galaxies (Kobulnicky \& Skillman, 1997, and references therein). \\

These results 
outline the problem of the dispersal and mixing of the newly synthesized 
elements in a starburst. Indeed, 
we know that there are some very massive evolved stars and one 
should expect the metals they have ejected 
to give some differences in the abundances between locations close and far 
from the star forming region (Kunth \& Sargent, 1986), which is not observed.
Therefore the question of the origin of the observed  heavy elements and the
 fate of the newly synthesized ones arise and will be discussed below.

%----------------------------------
\section{The mixing problem in Starbursts}

\begin{figure}
% psfile=#1 vsize=#2 angle=#3 hscale=#4 vscale=#5 hoffset=#6 voffset=#7
\plotfiddle{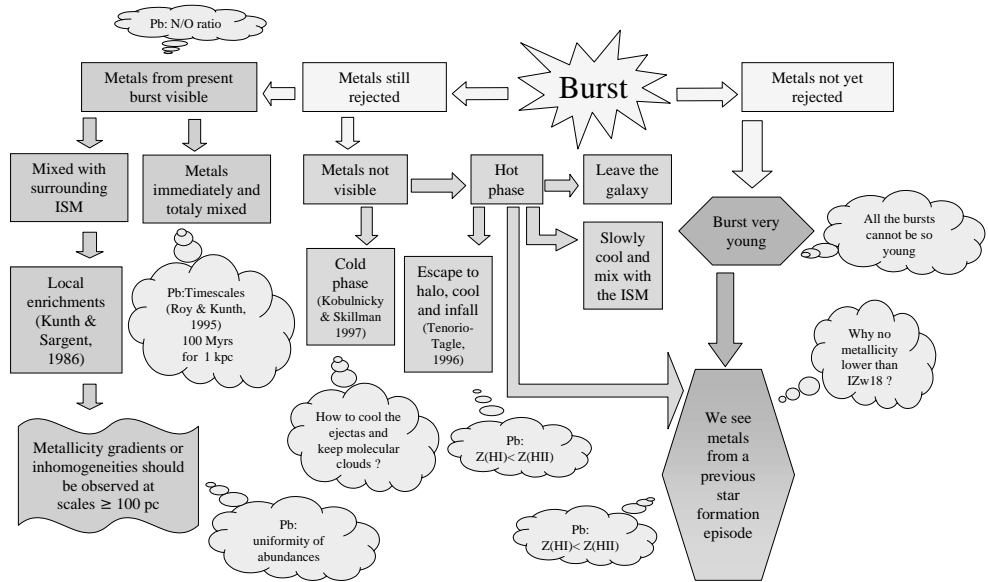}{74truemm}{0}{50}{50}{-200}{-43}
\caption{The mixing problem in Starbursts}
\label{mixing}
\end{figure}

The problem of the dispersal and mixing in starburst galaxies can be sketched
as in Fig. \ref{mixing}.

When a starburst galaxy is observed the metals produced by massive stars
may or may not be ejected into the interstellar medium. 
If metals have not yet been
ejected, it implies that the burst  is younger than 3 Myrs, which is not the
 case for
IZw 18 and for the majority of the starburst galaxies.
So, in most cases, 
the metals are in the process of being ejected. However these metals may still be either in visible stages or invisible.
If they are visible and produced by the ongoing burst the N/O 
ratio observed in starbursts (Masegosa et al, 1994, and references therein) 
would require a primary origin for the nitrogen, which is a strong constraint
 for standard
stellar evolution models.

The metals released should be dispersed and mixed with the interstellar 
medium surrounding the burst region as suggested by Kunth \& Sargent (1986). 
The size for the enriched regions should then be of the order of 
100 pc and some  discontinuities in the abundances should be observed at this 
scale which is not the case (Kobulnicky \& Skillman, 1997 and references 
therein). 

To explain the observed uniformity of abundances in starburst galaxies, an 
immediate and total mixing within the galaxies can be 
invoked but the timescale for such a mixing is around 100 Myrs for scales of
1 kpc (Roy \& Kunth, 1995), which is much larger than the derived age for bursts in IZw 18 and in many other starbursts.

Alternatively, metals ejected by the massive stars of
the present burst can be invisible because they are in a phase which do not
emit optical lines. They could be locked into a cold molecular phase, but it
 seems difficult to maintain dust or molecular clouds for a long time
 in a starburst
environment (Kobulnicky \& Skillman, 1997). 
The most probable possibility should be that the heavy elements ejected in 
the current burst of star formation are hidden in a hot $10^{6}$ K phase only 
visible in X-rays as suggested by Devost et al. (1997). In this case, the 
metals should follow a long excursion in the galactic halo before cooling 
and coming back into the central galactic region as suggested by Tenorio-Tagle 
(1996) or should be ejected and leave the galaxy or simply be dispersed in 
the interstellar medium where they will slowly cool and mix with the ambient 
medium. 
In all these cases, the observed metals must have been produced during
 a previous star formation event. In this case, 
two new questions arise: Why no galaxy with a metallicity lower than 
that of IZw 18 is found, and why do the abundance in the HI halo seem to be 
lower than the abundance in the HII regions ?

An explanation for the lack of starburst galaxies with a metallicity 
lower than the one of IZw 18 is that they simply do
not exist! If true this would mean that today galaxies have been pre-enriched
to this minimum metallicity. This would be a case for a pre-enrichement by a population III stars, unless they have been enriched via
a continuous star formation process at very low rate  since 
the epoch of formation of galaxies. Note also that one single previous burst analog to the current one is enough in most cases.

The second question is why the abundance in the HI halo seems to be lower 
than the abundance in the HII region ? If metals observed now have
been produced during a previous star formation event, one should expect that since
this epoch, the ejecta had had enough time to be dispersed and mixed over
the whole galaxy. Nevertheless, the results of Kunth et al (1994) concerning 
IZw 18 remains controversial (Pettini \& Lippman, 1995; van Zee, AJ, 1997, submitted) 
but new observations of other starbursts seem to indicate a trend for lower
abundances in the HI gas (Kunth et al, 1997, A\&A submitted; Thuan \& Izotov, 1997).
Therefore this crucial question is far from being solved and remains of major
 interest for our understanding of the mixing processes in the interstellar 
medium and of galaxiy evolution.

%----------------------------------
% TO ENCAPSULATE ONE FIGURE WITH FINE-TUNING
%\begin{figure}
% psfile=#1 vsize=#2 angle=#3 hscale=#4 vscale=#5 hoffset=#6 voffset=#7
%\plotfiddle{lastname_fig1.ps}{65truemm}{0}{50}{50}{-200}{-80}
%\caption{Caption of Fig.~1}
%\end{figure}

%----------------------------------
% TO ENCAPSULATE TWO FIGURES SIDE BY SIDE
%\begin{figure}
%\plottwo{lastname_fig2a.ps}{lastname_fig2b.ps}
%\caption{Caption of Figs.~2a and 2b}
%\end{figure}

%----------------------------------
% TO RESERVE APPROPRIATE SPACE FOR NON-ENCAPSULATED POSTSCRIPT FIGURES
%\begin{figure}
%\vspace{65truemm}
%\caption{Caption of Fig.~3}
%\end{figure}

%----------------------------------------------------------------------
\section{Conclusion}
We found WC stars in the metal poor galaxy IZw 18. It should imply
that mass loss rates at very low metallicity are more than twice the standard 
rate or that the binary channel of WR star formation is an important 
process. This indicates that metals are ejected in the ISM, but we found
no trace of inhomogeneities within the HII region which
should arise from these ejecta. This then indicate that the
metals ejected by the present massive stars are not seen because they are 
in a hot phase. This implies that the observed metals have been produced 
in a previous star formation event.
%----------------------------------------------------------------------
\acknowledgments 
I am indebted to Daniel Kunth and J.R. Roy for comments on this paper.
%----------------------------------------------------------------------

%----------------------------------------------------------------------

\begin{references}\small
\reference Devost D., Roy J-R., Drissen L., 1997, ApJ, 482, 765
\reference Izotov Y.I., Foltz C.B., Green R.F., Guseva N.G., Thuan T.X., 1997, ApJ, 487, L37
\reference Kobulnicky \& Skillman, 1997, ApJ, 489, 636
\reference Kunth D., \& Sargent W.L.W., 1986, ApJ, 300, 496
\reference Kunth D., Lequeux J., Sargent W.L.W., Viallefond F., 1994, A\&A, 282, 709
\reference Legrand F., Kunth D., Roy J-R., Mas-Hesse J.M., Walsh J., 1997, A\&A, 326, L17
\reference Masegosa J., Moles M., Campos-Aguilar A., 1994, ApJ, 420, 576
\reference Pettini M., \& Lippman K., 1995, A\&A, 297, L63
\reference Roy \& Kunth, 1995, A\&A, 294, 432
\reference Schaerer D., 1996, ApJ, 467, L17
\reference Tenorio-Tagle G., 1996, AJ, 111, 1641
\reference Thuan \& Izotov, 1997, ApJ, 489, 623

\end{references}
\end{document}